\documentclass[a4paper,useAMS,usenatbib,10pt]{nature}
\usepackage[dvips]{graphicx}
\usepackage{amsmath,amssymb}
\title{An upper limit on the contribution of accreting white dwarfs to the type Ia supernova rate}
\author{Marat Gilfanov$^{1,2}$ \& \'Akos Bogd\'an$^1$}
\date{}

\begin{document}
\maketitle

\begin{affiliations}
\item Max Planck Institut f\"ur Astrophysik, Karl-Schwarzschild-Str.1, 85741 
Garching, Germany 
\item Space Research Institute, Profsoyuznaya 
84/32, 117997 Moscow, Russia
\end{affiliations}

\begin{abstract}
\textbf{\boldmath
There is wide agreement that Type Ia supernovae (used as standard candles for cosmology) are associated with the thermonuclear explosions of  white dwarf stars.\cite{hillebrandt,livio_rev} The nuclear runaway that leads to the explosion   could start in a white dwarf gradually accumulating matter from a companion star until it reaches the Chandrasekhar limit,\cite{whelan} or could be triggered by the merger of two white dwarfs in a compact binary system.\cite{iben,webbink} The X-ray signatures of these two possible paths are very different. Whereas no strong electromagnetic emission is expected in the merger scenario until shortly before the supernova, the white dwarf accreting material from the normal star  becomes a source of copious X-rays  for $\sim 10^7$ yr before the explosion. This offers a means of determining which path dominates. Here we report that the observed X-ray flux from six nearby elliptical galaxies and galaxy bulges is a factor of $\sim 30-50$ less than predicted in the accretion scenario, based upon an estimate of the supernova rate  from their K-band luminosities. 
We conclude that no more than $\sim 5$ per cent of Type Ia supernovae in early type galaxies  can be produced by white dwarfs in accreting binary systems, unless their progenitors are much  younger than the bulk of the stellar population in these galaxies, or explosions of sub-Chandrasekhar white dwarfs make a significant contribution to the supernova rate.
}
\end{abstract}

The maximum possible mass of a carbon-oxygen white dwarf formed through standard stellar evolution can not not exceed $\approx 1.1-1.2M_\odot$.\cite{weidemann} 
Although  the nuclear detonation can start below the Chandrasekhar mass  ($\approx 1.38 M_\odot$), sub-Chandrasekhar models have so far failed to reproduce observed properties of Type Ia supernovae (SNIa),\cite{hoeflich96,nugent97} despite continuing effort.\cite{fink}
So the white dwarf needs to accreete $\Delta M\gtrsim 0.2M_\odot$ of matter before  the supernova explosion happens.

As accreted material accumulates on the white dwarf  surface, hydrogen shell burning is ignited, converting hydrogen to helium and, possibly, further to carbon and oxygen.  Depending on the mass accretion rate $\dot{M}$, it may proceed either in a (quasi-) steady regime or explosively,  giving rise to  Classical Nova events.\cite{nomoto07} Because mass is lost in Nova outbursts,\cite{prialnik} the white dwarf does not grow if  nuclear burning is unstable. For this reason the steady burning  regime is strongly preferred by the accretion scenario,\cite{livio_rev} limiting the range of the mass accretion rate 
relevant to the problem of SNIa progenitors to  $\dot{M}\gtrsim 10^{-7}~M_\odot$/yr.
In this regime  energy of hydrogen fusion is released in the form of electromagnetic radiation, with luminosity of $L_{WD,nuc}=\epsilon_H X \dot{M} \sim 10^{37}$ erg/s, where $\epsilon_H\approx 6\cdot 10^{18}$ erg/g  is energy release per unit mass and $X$ -- hydrogen mass fraction (the solar value of $X=0.72$ is assumed). 
The nuclear luminosity  exceeds by more than an order of magnitude the gravitational energy of accretion and maintains the effective temperature of the white dwarf  surface at the level (defined by the Stefan-Boltzmann law):  
\begin{equation}
T_{eff}\approx 45\ (\dot{M}/10^{-7} M_\odot/yr)^{1/4} (R_{WD}/10^{-2}R_\odot)^{-1/2}~ eV.  
\label{eq:teff}
\end{equation}
The black body
 spectrum of this temperature peaks in the soft X-ray band and, therefore, is prone to absorption by  interstellar gas and dust, especially at the lower end of the temperature range. 
Because  the white dwarf radius $R_{WD}$ decreases with its mass,\cite{panei} the  $T_{eff}$ increases as the white dwarf approaches the Chandrasekhar limit --  the signal, detectable at X-ray wavelengths, will be dominated  by the most massive white dwarfs.  Such sources are indeed observed in the Milky Way and nearby galaxies and are known as super-soft sources.\cite{sss} 

The Type Ia supernova rate $\dot{N}_{SNIa}$ scales with stellar mass and, hence, with near-infrared luminosity of the host galaxy.\cite{mannucci} The scale factor is calibrated through extensive observations of nearby galaxies and for E/S0 galaxies equals\cite{mannucci}  $\dot{N}_{SNIa}/L_K\approx 3.5\cdot 10^{-4}$ yr$^{-1}$ per $10^{10}~L_{K,\odot}$,  corresponding to one supernova in a few hundred years for a typical galaxy.
If the white dwarf mass increases at a rate $\dot{M}$,  a population of
\begin{equation}
N_{WD}\sim \frac{\Delta M}{\dot{M}\left<\Delta t\right>}\sim \frac{\Delta M}{\dot{M}}\dot{N}_{SNIa}
\label{eq:nwd}
\end{equation}
accreting white dwarfs is needed in order for one supernova to explode on average every $\left<\Delta t\right>=\dot{N}_{SNIa}^{-1}$ years (where $\Delta M$ is the difference between the Chandrasekhar mass and  the initial white dwarf mass).
With $\dot{M}\sim 10^{-7}-10^{-6}~M_\odot$/yr,   for a typical galaxy $N_{WD}\sim {\rm few} \times (10^2 -  10^3)$  -- the accretion scenario predicts a numerous population of accreting white dwarfs. The brightest and hottest of them may reveal themselves as super-soft sources,\cite{distefano} but the vast majority must remain unresolved or hidden from the observer, for example by   interstellar absorption. 
 The combined luminosity of this "sea" of accreting white dwarfs is 
\begin{equation}
L_{tot,nuc}=L_{WD,nuc}\times N_{WD}=\epsilon X\Delta M\dot{N}_{SNIa}
\label{eq:ltot}
\end{equation}
Unlike the number of sources, the luminosity  allows an accurate account for  absorption and bolometric corrections and therefore a quantitative comparison  with observations can be made.

We therefore collected archival data of X-ray (Chandra) and near-infrared (Spitzer and 2MASS) observations of several nearby gas-poor elliptical galaxies and for the bulge of M31 (Table \ref{tab:lx}).  Using K-band measurements to predict the SNIa rates, we computed combined X-ray luminosities of SNIa progenitors, based on a conservative, but plausible choice  of parameters: $\dot{M}=10^{-7}M_\odot$/yr and initial white dwarf mass of $1.2M_\odot$. The SNIa rate was reduced by a half in order to account for the fact that galaxies in our test-sample are somewhat older\cite{terlevich}  than those used to derive the rate.\cite{gallagher08} In computing the spectral energy distribution we took into account the dependence of the effective temperature on the white dwarf mass according to eq.(\ref{eq:teff}), and the effect of the  interstellar absorption (which  does not exceed a factor of $\sim 3-4$).  The X-ray and near-infrared data was prepared and analyzed as described elsewhere.\cite{bogdan08}  The observed X-ray luminosities were not corrected for absorption and include unresolved emission and emission from resolved compact sources with hardness ratio corresponding to $kT_{bb}\le 200$ eV.  The contribution of warm ionized gas was subtracted, when possible.  

Obviously, the observed values present upper limits on the luminosity of the hypothetical population of accreting white dwarf, as there may be other types of X-ray sources contributing to the observed emission.
As is clear from the Table \ref{tab:lx}, predicted luminosities surpass observed ones by a factor of $\sim 30-50$ demonstrating that the accretion scenario is inconsistent with observations by a large margin.

There exists a maximum rate at which hydrogen can burn on the white dwarf surface, $\dot{M}_{RG}\sim  10^{-6} M_\odot/$yr.\cite{nomoto07}  The excess material may leave the system in the form of a radiation driven wind\cite{hachisu} or may form a common envelope configuration.\cite{nomoto79, livio_rev} 
In both cases, because of the large photospheric radius, $\sim 10^2-10^3 $ R$_\odot$, the peak of the radiation is in the ultraviolet part of the spectrum and emission  from such an object will be virtually undetectable, due to interstellar absorption and dilution with the stellar light. 
However, there is a nearly universal consensus\cite{livio_rev}  that the common envelope configuration does not lead to the type Ia supernova explosion, producing a double white dwarf binary system instead.
 
In the wind regime, the white dwarf could grow in mass but it is a rather inefficient process because a significant  fraction of the transferred mass is lost in the wind.\cite{hachisu,vdh97}  Therefore a relatively massive, $M\gtrsim 1.3-1.7 M_\odot$, donor star is required in order for the white dwarf to reach the Chandrasekhar limit. As the lifetimes of such stars  do not exceed $\sim 2-5$ Gyrs, they may exist  only in the youngest of early type galaxies, in which no more than $\sim 30-40$ per cent of supernovae are detected.\cite{gallagher08}  We took this into account in our calculations  by halving the canonical value of the type Ia supernova rate\cite{mannucci}.
On a related note, in many elliptical galaxies small sub-populations of young stars are detected.\cite{schawinski} The ages of type Ia supernova progenitors are not very well constrained observationally, so it is possible in principle, that their progenitors are much younger than the bulk of the stellar population. However, given a small  fraction of young sub-populations in elliptical  galaxies (a few per cent or less), this would imply very high efficiency of young stars in producing supernovae and type Ia supernova rates  in spiral galaxies that are too high, much higher than observed.\cite{mannucci} This is therefore  not a likely scenario.

Thus, in early-type galaxies, white dwarfs accreting from a donor star in a binary system and detonating at the Chandrasekhar limit  do not contribute more than about 5 per cent to the observed type Ia supernova rate.
At present  the only viable alternative is the merger of two white dwarfs, so we conclude that type Ia supernovae in early-type galaxies arise predominantly from the double degenerate scenario. In late-type galaxies, in contrast, massive donor stars are available making the mass budget  less prohibitive, so that  white dwarfs can grow to the Chandrasekhar mass entirely  inside  an optically thick wind\cite{hachisu,vdh97} or via accretion of He-rich material from a He donor star.\cite{iben94} In addition, a star-forming environment is usually characterized by large amounts of neutral gas and dust, leading to increased absorption  obscuring  soft X-ray radiation from accreting white dwarfs. 
Therefore in late-type galaxies  the role of the accretion scenario may be significant.

\bigskip
{\it Acknowledgements}
The authors are grateful to  Friedrich Meyer and Hans Ritter for discussions of various aspects of mass transfer in binary systems and the role of the common envelope evolution and to Lev Yungelson for discussions of the SNIa progenitor problem in general.

This research has made use of Chandra archival data, provided by the Chandra X-ray Center (CXC) in the application package CIAO.  This research has made use of data products from Two Micron All 
Sky Survey, which is a joint project of the University of Massachusetts and 
the Infrared Processing and Analysis Center/California Institute of 
Technology, funded by NASA and the NSF. The Spitzer Space telescope is 
operated by the Jet Propulsion Laboratory, California Institute of 
Technology, under contract with NASA.

\noindent {\bf Author contributions:} The authors have contributed equally to this Letter.\\
{\bf Author infromation:} Correspondence and requests for materials should be addressed to M.G. (gilfanov@mpa-garching.mpg.de)

\begin{table} 
\caption{Comparison of the accretion scenario with observations.
Listed for each galaxy are:  name,  K-band luminosity, number of accreting white dwarfs and  X-ray luminosities in the soft (0.3--0.7 keV) band. 
The statistical errors for observed X-ray luminosities range from 20\% (NGC 3377) to less than 7\%.  The columns marked "predicted" display total number and combined X-ray luminosity (absorption applied) of accreting white dwarfs in the galaxy predicted in case the single degenerate scenario would produce all SNeIa. They  were computed assuming $\dot{M}=10^{-7}~M_\odot$/yr and initial white dwarf mass of $1.2M_\odot$. The $N_{WD}$ drops by a factor of 10 for $\dot{M}=10^{-6}~M_\odot$/yr.}
\vspace{0.5cm}
\centering
\begin{tabular}{l c c c c}
\hline
Name~~~~~~~~~  	& $L_K$ [$L_{K,\odot}$] & ~~~$N_{WD}$~~~ &\multicolumn{2}{c}{$L_X$ [erg/s]} \\
& ~~observed~~ & ~~predicted~~ &~~observed~~ & ~~predicted~~ \\
\hline
M32  &   $8.5\cdot 10^{  8}$ &  $25$ &$1.5\cdot 10^{36}$ & $7.1\cdot 10^{37}$	\\ 
NGC3377  &   $2.0\cdot 10^{10}$ &  $5.8\cdot 10^2$  &$4.7\cdot 10^{37}$ & $2.7\cdot 10^{39}$ \\
M31 bulge   &   $3.7\cdot 10^{10}$ &  $1.1\cdot 10^3$ &$6.3\cdot 10^{37}$ & $2.3\cdot 10^{39}$	\\
M105   &   $4.1\cdot 10^{10}$ 	&  $1.2\cdot 10^3$&$8.3\cdot 10^{37}$ & $5.5\cdot 10^{39}$  \\
NGC4278  &   $5.5\cdot 10^{10}$ &  $1.6\cdot 10^3$ &$1.5\cdot 10^{38}$ & $7.6\cdot 10^{39}$ \\
NGC3585  &   $1.5\cdot 10^{11}$  &  $4.4\cdot 10^3$ &$3.8\cdot 10^{38}$ & $1.4\cdot 10^{40}$ \\
\hline
\end{tabular}
\label{tab:lx}
\end{table}

\end{document}